\documentstyle[graphics,epsfig,amsmath,amssymb,natbib,psfrag,color]{mn2e}

\newcommand{\be}{\begin{equation}}
\newcommand{\e}{\end{equation}}
\newcommand{\bear}{\begin{eqnarray}}
\newcommand{\ear}{\end{eqnarray}}

\def\HI{H~{\sc i}\,}
\def\xh1{x_{H~{\sc i}}}
\def\xi{x^{i}_{{\rm H~{\sc{i}}}}\,}
\def\xb{\bar{x}_{{\rm H~{\sc{i}}}}}
\def\Ph1{P_{{H~{\sc{i}}}}}
\def\eh1{\eta_{{H~{\sc{i}}}}}
\def\snr{{\rm SNR}}

\def\Pb{{P_{b}}}
\def\Tb{{T_{b}}}
\def\aap{AAP}
\def\apj{ApJ}

\def\mnras{MNRAS}

\def\pasa{Pub. Astro. Soc. Australia}
\def\k{{\bf k}}

\begin{document}
\title[Non-Gaussianity on 21-cm power spectrum]{The effect of 
non-Gaussianity on error predictions  for the Epoch of Reionization 
(EoR) 21-cm power spectrum}
\author[Mondal et al.]{Rajesh Mondal$^{1, 2}$\thanks{rm@phy.iitkgp.ernet.in},
Somnath Bharadwaj$^{1, 2}$\thanks{somnath@phy.iitkgp.ernet.in}, Suman Majumdar$^3$,
\newauthor Apurba Bera$^1$ and Ayan Acharyya$^1$
\\
$^1$Department of Physics, Indian Institute of Technology, Kharagpur - 721302, India\\
$^2$Centre for Theoretical Studies, Indian Institute of Technology, Kharagpur - 721302, India\\
$^3$Department of Astronomy \& Oskar Klein Centre, AlbaNova, 
Stockholm University, SE-106 91 Stockholm, Sweden}

\date{Accepted 2015 January 20. Received 2014 December 14; in original form 2014 September 15}
\maketitle

\begin{abstract}
The Epoch of Reionization (EoR) 21-cm signal is expected to become increasingly non-Gaussian
as reionization proceeds.  We have used semi-numerical simulations to
study how this affects the error predictions for the EoR 21-cm power
spectrum.  We expect $\snr=\sqrt{N_k}$ for a Gaussian random field where 
$N_k$  is the number of Fourier modes in each $k$ bin.  We find
that non-Gaussianity is  important at high $\snr$ where it 
imposes an upper limit $[\snr]_l$. For a fixed volume $V$, it is
not possible to achieve $\snr > [\snr]_l$ even if $N_k$ is
increased. The value of $[\snr]_l$ falls as reionization proceeds,
dropping from $\sim 500$ at $\xb = 0.8-0.9$ to $\sim 10$ at $\xb =
0.15 $ for a $[150.08\, {\rm Mpc}]^3$ simulation.  We show that 
it is possible to interpret $[\snr]_l$ in terms of the trispectrum, and 
we expect $[\snr]_l \propto \sqrt{V}$  if the volume is increased. 
For $\snr \ll [\snr]_l$ we find $\snr = \sqrt{N_k}/A $ with 
$A \sim 0.95 - 1.75$, roughly consistent with the Gaussian prediction. We
present a fitting formula for the $\snr$ as a function of $N_k$, with
two parameters $A$ and $[\snr]_l$ that have to be determined using
simulations.  Our results are relevant for predicting the sensitivity
of different instruments to measure the EoR 21-cm power spectrum,
which till date have been largely based on the Gaussian assumption.
\end{abstract}


\section{Introduction}
\label{sec:intro}
Observations of the redshifted 21-cm signal from neutral hydrogen
(\HI) are a very promising probe of the Epoch of Reionization (EoR),
and there is a considerable observational effort underway to detect
the EoR 21-cm power spectrum e.g.
{GMRT\footnote{http://www.gmrt.ncra.tifr.res.in}} \citep{paciga13},
{LOFAR\footnote{http://www.lofar.org/}} \citep{yatawatta13, haarlem13},
{MWA\footnote{http://www.haystack.mit.edu/ast/arrays/mwa/} \citep{tingay13, bowman13}}, and
{PAPER\footnote{http://eor.berkeley.edu/}} \citep{parsons2014, jacobs2014}. 
Observing the EoR 21-cm signal is one of the key
scientific goals of the future telescope
{SKA\footnote{http://www.skatelescope.org/} \citep{mellema13}}. It is
important to have quantitative predictions of both, the expected EoR
21-cm power spectrum and the sensitivity of the different instruments
to measure the expected signal.

On the theoretical and computational front, a considerable amount of
effort has been devoted to simulate the expected EoR 21-cm signal
(e.g.  \citealt{gnedin00, zahn05, mellema06a, trac07, thomas09,
  battaglia13}).  There also have been several works to quantify the
sensitivity to the EoR signal for different instruments
(e.g. \citealt{morales05, mcquinn06}). \citet{Beardsley13},
\citet{jensen13} and \citet{pober14} have recently
made quantitative predictions for detecting the EoR 21-cm power
spectrum with the MWA, LOFAR, SKA and PAPER respectively.

The sensitivity of any instrument to the EoR 21-cm power spectrum is
constrained by the errors, a part of which arises from the system
noise of the instrument and another component which is inherent to the
signal that is being detected (cosmic variance).  It is commonly
assumed, as in all the sensitivity estimates mentioned earlier, that
the system noise and the EoR 21-cm signal are both independent
Gaussian random variables. This is a reasonably good assumption at
large scales in the early stages of reionization when the \HI is
expected to trace the dark matter.  Ionized bubbles, however,
introduce non-Gaussianity \citep{bharadwaj05} and the 21-cm signal is
expected to become highly non-Gaussian as the reionization
proceeds.  This transition in the 21-cm signal is clearly visible in
Figure \ref{fig:HI_map}.

In this {\em Letter} we use semi-numerical simulations of the EoR 21-cm signal to
study the effect of non-Gaussianities on the error estimates for the
21-cm power spectrum.  Not only is this important for correctly
predicting the sensitivity of the different instruments, it is also
important for correctly interpreting the observation once an actual
detection has been made.  The entire analysis here focuses on the
errors which are intrinsic to the 21-cm signal, and we do not consider
the system noise corresponding to any particular instrument.

Throughout the {\em Letter}, we have used the Planck+WP best fit values of
cosmological parameters $\Omega_{m0}=0.3183$, $\Omega_{\Lambda0}=0.6817$
, $\Omega_{b0}h^2=0.022032$, $h=0.6704$
, $\sigma_8=0.8347$, and $n_s=0.9619$ \citep{planck13}.

\section{Simulating The 21-cm Maps}
\label{sec:simulation}
The evolution of the mass averaged neutral fraction $\xb(z)$ during
EoR is largely unconstrained. Instead of choosing a particular model
for $\xb(z)$, we have fixed the redshift $z=8$ and considered
different values of $\xb$ at an interval of $0.1$ in the range $1.0
\ge \xb \ge 0.3$ in addition to $\xb =0.15$. For each value of $\xb$
we have simulated $21$ statistically independent realizations of the
21-cm map which were used to estimate the mean $\Pb(k)$ and the rms.
fluctuation (error) $\delta \Pb(k)$ of the 21-cm power spectrum. We
have used these to study how $\Pb(k)$ and particularly $\delta \Pb(k)$
evolve as reionization proceeds {\it i.e.}  $\xb$ decreases.

The simulations are based on three main steps: (1.) determine the dark
matter distribution at the desired redshift, (2.) identify the
collapsed halos (3.) generate the reionization map using an excursion
set formalism \citep{furlanetto04} under the assumption that the
collapsed halos host the ionizing sources and the hydrogen exactly
traces the dark matter.

We have used a particle-mesh N-body code to simulate the $z=8$ dark
matter distribution in a $V_1=[150.08\, {\rm Mpc}]^3$ comoving volume with
a $2144^3$ grid using $1072^3$ dark matter particles.  The standard
Friends-of-Friends (FoF) algorithm was used to identify collapsed dark
matter halos from the output of the N-body simulation. We have used a
fixed linking length $0.2$ times the mean inter-particle separation,
and require a halo to have at least $10$ particles which corresponds to
a minimum halo mass of $7.3 \times10^8 h^{-1} M_{\odot}$.
 
We have assumed that the number of ionizing photons from a collapsed
halo is proportional to its mass. It is possible to achieve different
values of $\xb$ by appropriately choosing this proportionality factor.
The ionizing photon field was used to construct the hydrogen
ionization fraction and the \HI distribution using the homogeneous
recombination scheme of \citet{choudhury09}.  Following
\citet{majumdar13}, the simulated \HI distribution was mapped to
redshift space to generate the 21-cm maps. The steps outlined in this
paragraph used a low resolution grid $8$ times coarser than the N-body
simulations.

\begin{figure}
\centering
\includegraphics[width=.489\textwidth, angle=0]{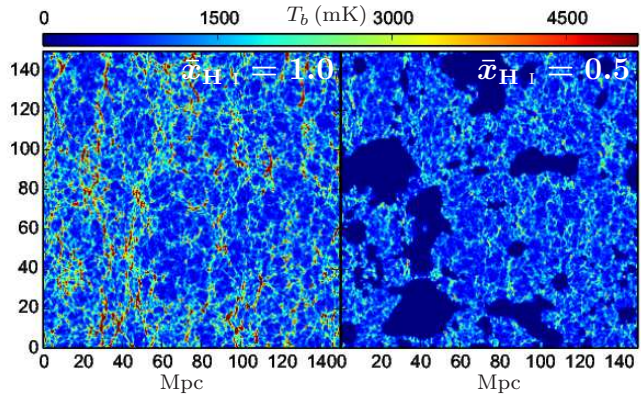}
\put(-185.1, -2.5){${\rm Mpc}$}
\put(-65.1, -2.5){${\rm Mpc}$}
\put(-137, 137){$T_b\, ({\rm mK})$}
\put(-177, 115){\Large \textcolor{white}{\boldmath$\xb=1.0$}}
\put(-65, 115){\Large \textcolor{white}{\boldmath$\xb=0.5$}}
\caption{A section through one of the  simulated redshift  space
\HI brightness temperature maps for $\xb=1.0$  (left) which is 
largely a Gaussian random field, and $\xb=0.5$ (right) which 
has considerable non-Gaussianity due to the discrete ionized 
bubbles visible in the image. The redshift space distortion is 
with  respect to a distant observer located along the horizontal 
axis.}
\label{fig:HI_map}
\end{figure}

Figure \ref{fig:HI_map} shows a section through one of the simulated
three dimensional 21-cm maps with $\xb=1$ and $0.5$ in the left and
right panels respectively. The brightness temperature ${\rm T_b}({\bf x})$
is to a good approximation a Gaussian random field for $\xb=1$. The
homogeneous recombination scheme implemented here predicts an
``inside-out'' reionization where the high density regions are ionized
first and the low density regions later. The image at $\xb=0.5$ is
dominated by several ionized bubbles which preferentially mask out the
high density regions, the low density regions are left untouched. We
expect the statistics of ${\rm T}_b({\bf x})$, or equivalently
$\tilde{{\rm T}}_b({\bf k})$ its Fourier transform, to show considerable
deviations from the original Gaussian distribution. The induced
non-Gaussianity will reflect in the sizes and distribution of the
ionized bubbles and we expect the non-Gaussianity to increase as
reionization proceeds.

\section{Results}
\label{sec:result}
\begin{figure}
\psfrag{pk}[c][c][1][0]{${\Delta^2_{b}}\, \, ({\rm mK^2})$}
\psfrag{k}[c][c][1][0]{$k\, \, ({\rm Mpc}^{-1}$)}
\psfrag{xHI=1.0}[c][c][1][0]{$\xb$= $1.0$\, }
\psfrag{0.8}[c][c][1][0]{$0.8$}
\psfrag{0.6}[c][c][1][0]{$0.6$}
\psfrag{0.5}[c][c][1][0]{$0.5$}
\psfrag{0.3}[c][c][1][0]{$0.3$}
\psfrag{10}[c][c][1][0]{$10$}

\centering
\includegraphics[width=.339\textwidth, angle=-90]{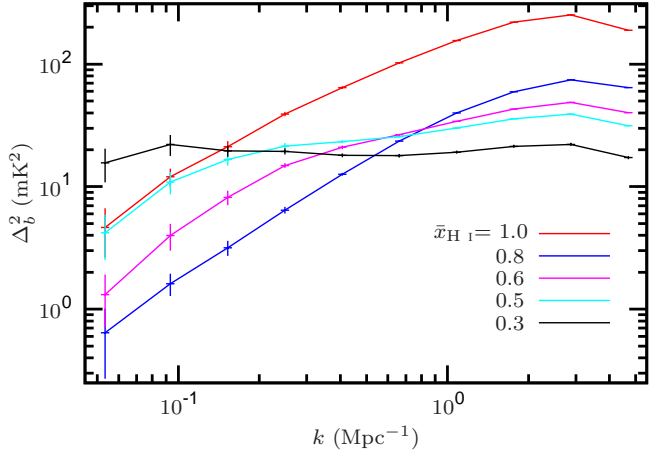}
\caption{The mean squared 21-cm brightness temperature 
fluctuations $\Delta_b^2(k)$ and its $1-\sigma$ error bars
for the $\xb$ values shown in the figure.}
\label{fig:pk_HI}
\end{figure}

\begin{figure}
\psfrag{snr}[c][c][1][0]{{$\snr$}}
\psfrag{sqrtnk}[c][c][1][0]{{$\sqrt{N_k}$}}
\psfrag{k}[c][c][1][0]{$k\, \, ({\rm Mpc}^{-1})$}
\psfrag{Initial}[c][c][1][0]{Initial}
\psfrag{xhi0.8}[c][c][1][0]{$\xb$= $0.8$\, \, \, }
\psfrag{0.6}[c][c][1][0]{$0.6$}
\psfrag{0.4}[c][c][1][0]{$0.4$}
\psfrag{0.3}[c][c][1][0]{$0.3$}
\psfrag{0.15}[c][c][1][0]{$0.15$}
\psfrag{10}[c][c][1][0]{$10$}

\psfrag{0.05}[c][c][1][0]{$0.05$}
\psfrag{0.09}[c][c][1][0]{$0.09$}
\psfrag{0.25}[c][c][1][0]{$0.25$}
\psfrag{0.25}[c][c][1][0]{$0.25$}
\psfrag{0.41}[c][c][1][0]{$0.41$}
\psfrag{0.66}[c][c][1][0]{$0.66$}
\psfrag{1.08}[c][c][1][0]{$1.08$}
\psfrag{1.76}[c][c][1][0]{$1.76$}
\psfrag{2.88}[c][c][1][0]{$2.88$}
\psfrag{4.69}[c][c][1][0]{$4.69$}

\centering
\includegraphics[width=.381\textwidth, angle=-90]{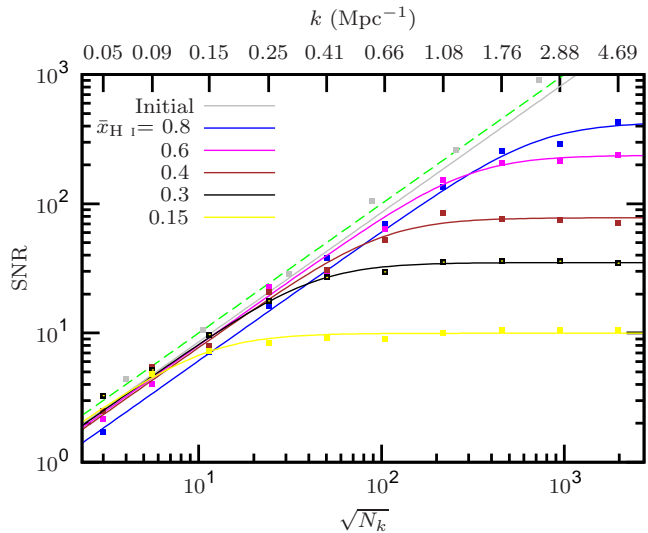}
\caption{This shows the $\snr$ as a function of $\sqrt{N_k}$.  The
  $45^{\circ}$ dashed line shows the $\snr$ expected for a Gaussian
  random field, and `Initial' refers to the input linear density
  fluctuations used for the dark matter N-body simulations. For the
  $\xb$ values mentioned in the figure, the data points (squares) show
  the simulated $\snr$ and the solid lines show the fit given by
  eq. (\ref{eq:a1}). We have used $10$ equally spaced logarithmic
  bins, and the $k$ value corresponding to each bin is shown in the
  top $x$ axis.}
\label{fig:delta_pk}
\end{figure}

Figure \ref{fig:pk_HI} shows the brightness temperature fluctuation
${\Delta^2_{b}}(k)=k^3 \Pb (k)/2\pi^2$ as a function of $k$ for
different values of $\xb$. The average power spectrum $\Pb (k)$ and
the $1-\sigma$ errors $ \delta \Pb (k)$ were calculated using 21
independent realizations of the simulation, and the $k$ range has been
divided into $10$ equally spaced logarithmic bins.  Note the change in
${\Delta^2_{b}}(k)$ as reionization proceeds.  At $\xb \sim 0.5$, the
non-Gaussian Poisson noise of the discrete ionized regions makes a
considerable contribution to ${\Delta^2_{b}}(k)$ at length-scales that
are larger than the typical bubble radius.  The ionized regions
percolate at smaller $\xb$ where the Poisson noise of the surviving
discrete \HI regions makes a considerable contribution to
${\Delta^2_{b}}(k)$.  While these effects have an imprint on the
predicted ${\Delta^2_{b}}(k)$, the power spectrum does not capture the
fact that the predicted signal is non-Gaussian. The error estimates
for the power spectrum, however, are affected by the non-Gaussianity
of the 21-cm signal.

We expect the signal to noise ratio to follow $\snr=\Pb (k)/\delta \Pb
(k)=\sqrt{N_k}$ if the 21-cm signal is a Gaussian random field, $N_k$
here is the number of Fourier modes in each $k$ bin. We have tested
the Gaussian assumption by plotting the simulated $\snr$ as a function
of $\sqrt{N_k}$ in Figure \ref{fig:delta_pk} where the $45^{\circ}$
dashed line shows the values expected for a Gaussian random field. We
see that the input linear power spectrum used in the dark matter
N-body simulations follows this over the entire range.  In contrast,
the $\snr$ for the 21-cm power spectrum shows a different
behaviour. For $\xb \ge 0.3$ we find the expected $\snr \propto
\sqrt{N_k}$ behaviour at $\snr \le 10$, however the $\snr$ values are
$0.95-1.75$ times than those predicted for a Gaussian random
field.  For larger $\snr$ it increases slower than $\sqrt{N_k}$ and
finally saturates at a limiting value $[\snr]_l$. The limiting value
$[\snr]_l$ decreases as reionization proceeds ($\xb$ falls).  We do
not explicitly see the $\snr \propto \sqrt{N_k}$ behaviour for $\xb
=0.15$, this possibly exists at $\snr <1$ which is outside the range
that we have considered. In this case the $\snr$ values are close to
$[\snr]_l$ for the entire range that we have considered.

\begin{figure}
\psfrag{snr}[c][c][1][0]{{$\snr$}}
\psfrag{sqrtnk}[c][c][1][0]{{$\sqrt{N_k}$}}
\psfrag{bins=10}[c][c][1][0]{Bins $=10$\, \,}
\psfrag{20}[c][c][1][0]{$20$}
\psfrag{40}[c][c][1][0]{$40$}
\psfrag{10}[c][c][1][0]{$10$}

\centering
\includegraphics[width=.339\textwidth, angle=-90]{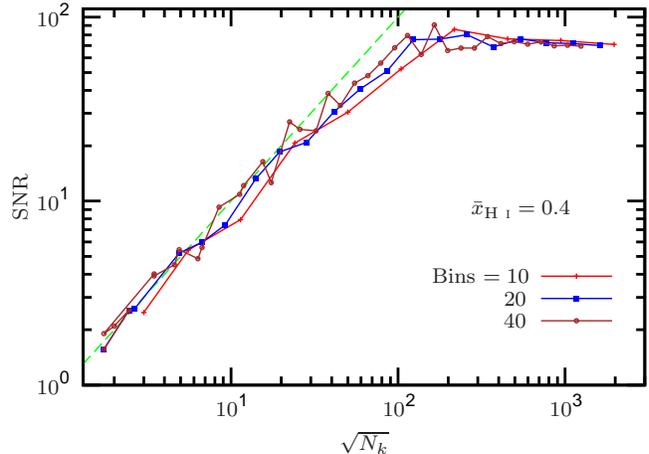}
\put(-69.1, -81.5){$\xb=0.4$}
\caption{For $\xb=0.4$, this shows the simulated $\snr$ as a function of
$\sqrt{N_k}$ for $10$, $20$ and $40$ equally spaced logarithmic bins.}
\label{fig:snr_bin}
\end{figure}

The equally spaced logarithmic bins that we have used imply a relation
$N_k= C k^3$ between $N_k$ and $k$    (where $C$ is a 
constant), and the corresponding $k$ values are shown on the upper
$x$ axis of Figure \ref{fig:delta_pk}. It is therefore plausible that,
in addition to $\xb$, the deviations from the Gaussian predictions may
also depend on $k$. To test this we have also considered $20$ and $40$
equally spaced logarithmic bins (Figure \ref{fig:snr_bin}).  The
relation between $N_k$ and $k$ changes ({\it i.e.} 
the value of $C$ changes) if we change the number of bins, however
we find that curves showing the $\snr$ as a function of $N_k$ do not
change.  We therefore conclude that the effect of the non-Gaussianity 
on the $\snr$ (or equivalently  $\delta P_b(k)/P_b(k)$) does not depend 
on $k$,  it   depends only on $\xb$ and $N_k$.

\begin{figure}
\psfrag{snrlimit}[c][c][1][0]{{$[\snr]_l$}}
\psfrag{xhi}[c][c][1][0]{{$\xb$}}
\psfrag{a}[c][c][1][0]{$A$}
\psfrag{bins=10}[c][c][1][0]{Bins $=10$\, \, \,}
\psfrag{20}[c][c][1][0]{$20$}

\centering
\includegraphics[width=.3\textwidth, angle=-90]{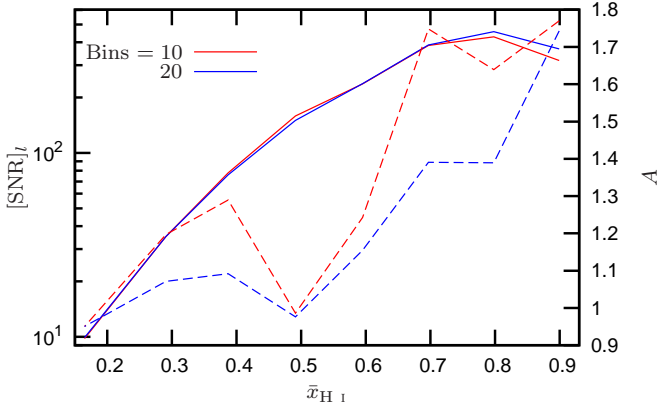}
\caption{This shows how the best fit parameters $A$ (dashed) and 
$[\snr]_l$ (solid) vary with $\xb$. The results are shown for $10$ 
and $20$ equally spaced logarithmic bins.}
\label{fig:fit_p}
\end{figure}

We find that the function 
\begin{equation}
\snr=\frac{\sqrt{N_k}}{A} \left[ 1+\frac{N_k}{(A [\snr]_l)^2} \right]^{-0.5} 
\label{eq:a1}
\end{equation}
provides a good fit to the simulated $\snr$. For each value of $\xb$,
we have used a least-square fit to obtained the best fit $A$ and
$[\snr]_l$.  The solid curves in Figure \ref{fig:delta_pk} show the
fit to the $\snr$ given by eq. (\ref{eq:a1}) using the best fit
parameters.  Figure \ref{fig:fit_p} shows the best fit parameters $A$
and $[\snr]_l$ as a function of $\xb$.  The parameter $A$ quantifies
the deviation from the Gaussian prediction in the low $\snr$ regime
($\snr \ll [\snr]_l$) where we have $\snr = \sqrt{N_k}/A$.  
We find that the value of $A$ increases from $A \sim 0.95$ 
at $\xb =0.15$ to $A \sim 1.75$ at $\xb =0.9$. 
Surprisingly, in this regime the $\snr$
approaches the Gaussian prediction as the reionization proceeds.  In
contrast, the value of $[\snr]_l$ decreases by a factor of $\sim 50$
as the $\xb$ falls from $0.9$ to $0.15$. The deviations from the
Gaussian predictions seen at large $\snr$ increase as reionization
proceeds.

\section{Modelling the $\snr$.}
The power spectrum $\Pb(k)$ and the trispectrum $\Tb(\k_1,\k_2,\k_3,\k_4)$
of the brightness temperature fluctuations $\tilde{{\rm T}}_b(a) \equiv 
\tilde{{\rm T}}_b({\bf k_a})$
are respectively defined through 
\begin{equation}
\langle \tilde{{\rm T}}_b({ a}) \tilde{{\rm T}}_b(b) \rangle =
V \delta_{a+b,0}  \Pb(a)
\end{equation}
and 
\begin{eqnarray*}
\langle \tilde{{\rm T}}_b(a) \tilde{{\rm T}}_b(b)
\tilde{{\rm T}}_b(c) \tilde{{\rm T}}_b(d) \rangle &=&
V^2  [ \, \delta_{a+b,0} \,  \delta_{c+d,0} \, \Pb(a) \Pb(c)
\nonumber \\
+ \delta_{a+c,0}  \delta_{b+d,0}  \Pb(a) \Pb(b)
&+&\delta_{a+d,0}  \delta_{b+c,0}  \Pb(a) \Pb(b)] 
\nonumber \\
&+& V \delta_{a + b + c + d,0}  \, \Tb(a,b,c,d)
\end{eqnarray*}
where $V$ refers to the comoving volume of the region under
consideration and $\langle ... \rangle$ denotes an ensemble average
over different realizations of the fluctuations. We use these  
to calculate the mean and the variance of the binned power spectrum estimator 
which we define as 
\begin{equation}
\hat{\Pb}(k)= (N_k V)^{-1} \sum_{a}  \tilde{{\rm T}}_b(a) 
\tilde{{\rm T}}_b(-a) \, ,
\label{eq:est}
\end{equation}
where the sum $\sum_a$  extends over all the  Fourier modes $\k_a$ within the bin,
and $k$ is the representative comoving wave number for the bin. The bins 
here are spherical shells of width $\Delta k$ (which varies from bin to bin).
The modes $\k_a$ and $-\k_a$
do not give independent estimates of the power spectrum. We restrict 
the sum $\sum_a$ to half the spherical shell, and $N_k$  refers to the number 
of Fourier modes in this volume.  
 
We then have 
\begin{equation}
\langle \hat{\Pb}(k) \rangle  = \bar{\Pb}(k) =(N_k)^{-1} \sum_a \Pb(a)
\end{equation}
which is the bin averaged power spectrum, and the variance 
\begin{equation}
 \langle [\delta \hat{\Pb}(k)]^2  \rangle =
 [\delta {\Pb}(k)]^2 =(N_k)^{-1} 
 \overline{\Pb^2}(k) + V^{-1} \bar{\Tb}(k,k)
\end{equation}
where 
\begin{equation}
 \overline{\Pb^2}(k)=(N_k)^{-1} \sum_a \Pb^2(a) 
\end{equation}
and 
\begin{equation}
 \bar{\Tb}(k,k)=(N_k)^{-2} \sum_{a,b}  \Tb(a,-a,b,-b)
\end{equation}
are the square of the power spectrum and the trispectrum 
respectively  averaged over the bin.  

The $\snr \equiv  \bar{\Pb}(k)/[\delta {\Pb}(k)]$ can be cast 
in the form of eq. (\ref{eq:a1}) provided we identify 
\begin{equation}
A = \sqrt{\frac{\overline{\Pb^2}(k)}{[\bar{\Pb}(k)]^2}}
\label{eq:A1}
\end{equation}
and 
\begin{equation}
[\snr]_l=\sqrt{\frac{[\bar{\Pb}(k)]^2 V}{\bar{\Tb}(k,k)}}
\label{eq:A2}
\end{equation}

Our calculation (eq. \ref{eq:A1}) shows that   $A$  arises from the fact
that the power spectrum varies across the different Fourier modes which
contribute to a  bin. This implies that $\overline{\Pb^2}(k) > [\bar{\Pb}(k)]^2$ 
whereby $A >1$  even for a purely Gaussian random field. This explains why the 
$\snr$ values for the `Initial' Gaussian random field (Figure \ref{fig:delta_pk})  
are lower than those predicted by $\snr=\sqrt{N_k}$, even though both have the same 
slope.

\begin{figure}
\psfrag{snr}[c][c][1][0]{{$\snr$}}
\psfrag{sqrtnk}[c][c][1][0]{{$\sqrt{N_k}$}}
\psfrag{150mpc}[c][c][1][0]{$(150{\rm Mpc})^3$\, \,}
\psfrag{215mpc}[c][c][1][0]{$(215{\rm Mpc})^3$\, \,}
\psfrag{10}[c][c][1][0]{$10$}

\centering
\includegraphics[width=.339\textwidth, angle=-90]{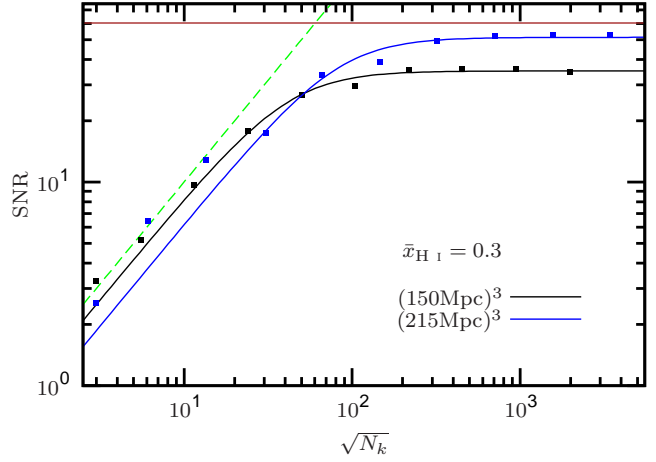}
\put(-95.1, -99.5){$\xb=0.3$}
\caption{This shows a comparison of the  results from 
simulations with  two  different box sizes, for a fixed  
$\xb=0.3$. The solid curves show eq. (\ref{eq:a1}) with  the 
best fit parameters for the  respective data points (squares).  
The horizontal solid line shows $[\snr]_l$ for the smaller 
volume scaled by the factor $\sqrt{V_2/V_1}$, and the dashed 
line shows  $\snr=\sqrt{N_k}$.}
\label{fig:snr_comp}
\end{figure}

The limiting $\snr$ (eq. \ref{eq:A2}) depends on the trispectrum. 
Considering a  Gaussian random field first, the statistics  are completely
specified by the power spectrum
and   $\Tb(\k_1,\k_2,\k_3,\k_4)=0$. We  expect $\snr=\sqrt{N_k}/A$ to hold throughout
in this case. However, the brightness temperature fluctuations become increasingly  non-Gaussian 
as  reionization proceeds. We expect a non-zero  trispectrum to develop 
and increase as reionization proceeds. This is borne out by Figure~\ref{fig:fit_p} where 
$[\snr]_l$ is found to fall  as $\xb$ declines. Our results (Figure \ref{fig:snr_bin}) also indicate 
that the ratio $[\bar{\Pb}(k)]^2/\bar{\Tb}(k,k) \propto [\snr]_l^2$ is roughly independent of $k$, at least 
over the $k$ range accessible through rebinning the data in the figure.  Finally, eq. (\ref{eq:A2}) implies
that we expect the limiting $\snr$ to scale as $[\snr]_l \propto \sqrt{V}$ with the  simulation volume. 

We have investigated  the volume dependence of the $\snr$ by carrying out $21$ independent realizations 
of  a larger simulation with comoving volume $V_2=[215 \, {\rm Mpc}]^3$, maintaining the same spatial 
resolution   as in Section \ref{sec:simulation}. 
Figure \ref{fig:snr_comp} shows a comparison of the $\snr$ between the 
smaller ($V_1$) and the larger ($V_2$)
simulations for $\xb=0.3$. In both cases the power spectrum was 
evaluated in $10$ equally spaced logarithmic bins. Note that the Fourier
modes, $N_k$ and $k$  corresponding to the $10$  bins are different in 
the two sets of simulations which are being compared.  The results 
for the larger simulation are qualitatively similar to those for the 
smaller one, though the $\snr$ values are different. 
We find that the functional form given by eq. (\ref{eq:a1}) does not 
provide  a very good fit  at small $N_k$ for the larger simulation. 
This is possibly because the value of $A$ (eq. \ref{eq:A1}) varies from
bin to bin. The fit, however, is very good  at large $N_k$ where the behaviour 
is  dominated by $[\snr]_l$.  The horizontal solid line in the figure   shows $[\snr]_l$ 
for the smaller simulation scaled   by the factor $\sqrt{V_2/V_1}$.  We find that $[\snr]_l$ calculated 
from the larger simulation is roughly consistent with this solid line. This  validates  the 
 $[\snr]_l \propto \sqrt{V}$  dependence  predicted  by eq. (\ref{eq:A2}). 
We find similar results for the other values of $\xb$ not shown here.

\section{Discussion and Conclusions}

We may think of the EoR 21-cm signal as a combination of two
components, one a Gaussian random field and another a non-Gaussian
component from the discrete ionized bubbles. The picture is slightly
changed as the reionization proceeds and the ionized regions
percolate. The non-Gaussian component then arises from the discrete \HI
clumps.  The Gaussian components in the different Fourier modes
$\tilde{\rm T}_b({\k})$ are independent, the non-Gaussian components
however are  correlated - this being quantified through the bispectrum 
\citep{bharadwaj05}, trispectrum etc.
 The contribution to $\delta \Pb (k)/ \Pb
(k)$ from the Gaussian component comes down as $1/\sqrt{N_k}$, whereas
the non-Gaussian contribution remains fixed even if $N_k$ is
increased.  The Gaussian assumption gives a reasonable description at
low $\snr$, the non-Gaussian contribution however sets an upper limit
$[\snr]_l$. For a fixed volume $V$, 
it is not possible to increase the $\snr$ beyond $[\snr]_l$  by
combining the signal from more Fourier modes.  The non-Gaussianity
increases as reionization proceeds, and $[\snr]_l$ falls from $\sim
500$ at $\xb = 0.8-0.9$ to $\sim 10$ at $\xb = 0.15 $ for the
$[150.08\, {\rm Mpc}]^3$ simulations. 

The limiting signal to noise ratio $[\snr]_l$ is proportional to 
$\sqrt{V}$, and it is possible to achieve a high  $\snr$ by increasing 
the volume. The value of $\sqrt{N_k}$ in eq. (\ref{eq:a1}), however, 
also scales as $\propto \sqrt{V}$  for  a fixed bin width $\Delta k$.
Although it is possible to increase the $\snr$ by increasing the volume,
the relative contribution from the non-Gaussianity   
($[\bar{\Pb}(k)]^2/\bar{\Tb}(k,k)$) does not vary with $V$.

In the present analysis we have used a simple model of reionization, 
and held $z=8$ fixed.
The predictions will be different if effects like inhomogeneous
recombination are included, and the evolution of $\xb$ with $z$ is
taken into account. The present work highlights the fact that
non-Gaussian effects could play an important role in  the error
predictions for the EoR 21-cm power spectrum. We plan to 
consider the implications for the different EOR experiments in future work.

\end{document}